\documentclass[twocolumn]{aastex62}

\usepackage{graphicx}
\usepackage{grffile}
\usepackage{natbib}
\usepackage{amsmath}
\usepackage{url}
\usepackage{enumitem}

\newcommand{\transpose}{\intercal}
\newcommand{\five}{NG5~}
\newcommand{\nine}{NG9~}

\newcommand{\epsr}{J1713$+$0747}

\newcommand{\an}{a_{n}}

\newcommand{\nchan}{n_{\textrm{\footnotesize chan}}}
\newcommand{\nbin}{n_{\textrm{\footnotesize bin}}}
\newcommand{\neig}{n_{\textrm{\footnotesize eig}}}
\newcommand{\nB}{n_{\textrm{\footnotesize B}}}
\newcommand{\nt}{n_{\footnotesize t}}
\newcommand{\nulo}{\nu_{\textrm{lo}}}
\newcommand{\nuhi}{\nu_{\textrm{hi}}}

\accepted{for publication in ApJ}

\begin{document}

\title{Frequency-Dependent Template Profiles for High Precision Pulsar Timing}
\shorttitle{Wideband Templates for Pulsar Timing}
\shortauthors{T.\,T.\,Pennucci}
\author[0000-0001-5465-2889]{Timothy T. Pennucci}
\affiliation{Hungarian Academy of Sciences MTA-ELTE ``Extragalatic Astrophysics'' Research Group, Institute of Physics,\\E\"{o}tv\"{o}s Lor\'{a}nd University, P\'{a}zm\'{a}ny P. s. 1/A, Budapest 1117, Hungary}

\email{tim.pennucci@nanograv.org}

\begin{abstract}
\makeatletter{}Pulsar timing experiments require high fidelity template profiles in order to minimize the biases in pulse time-of-arrival (TOA) measurements and their uncertainties.
Efforts to acquire more precise TOAs given fixed effective area of telescopes, finite receiver noise, and limited integration time have led pulsar astronomers to the solution of implementing ultra-wideband receivers.
This solution, however, has run up against the problem that pulse profile shapes evolve with frequency, which raises the question of how to properly measure and analyze TOAs obtained using template-matching methods.
This paper proposes a new method for one facet of this problem, that of template profile generation, and demonstrates it on the well-timed millisecond pulsar \epsr.
Specifically, we decompose pulse profile evolution into a linear combination of basis eigenvectors, the coefficients of which change slowly with frequency such that their evolution is modeled simply by a sum of low degree piecewise polynomial spline functions.
These noise-free, high fidelity, frequency-dependent templates can be used to make measurements of so-called ``wideband TOAs'' simultaneously with an estimate of the instantaneous dispersion measure.  The use of wideband TOAs is becoming important for pulsar timing array experiments, as the volume of datasets comprised of conventional, subbanded TOAs are quickly becoming unwieldly for the Bayesian analyses needed to uncover latent gravitational wave signals.
Although motivated by high precision timing experiments, our technique is applicable in more general pulsar observations.

\end{abstract}


\NewPageAfterKeywords

\section{Introduction}
\label{sec:introduction}
\makeatletter{}Pulsar timing stands out in the field of observational astrophysics for the level of precision to which it can measure physical quantities \citep{L&K05,Manchester17}.
Of current interest in the pulsar timing community is the possibility to uncover gravitational waves imprinted upon the pulsar signal \citep{Manchester13b,McLaughlin14,IPTA16a,IPTA16b}.
The deviations induced by gravitational waves in the timing of a pulsar are thought to be no larger than several to tens of nanoseconds \citep{Spolaor15}.
The high precision experiments attempting to measure these gravitational waves are referred to as pulsar timing array (PTA) experiments \citep{FB90}.

\subsection{Precision Timing, PTAs, \& Wideband Receivers}
\label{subsec:ptas}

The North American Nanohertz Observatory for Gravitational Waves (NANOGrav)\footnote{\url{nanograv.org}} \citep{McLaughlin13} is one of several ongoing PTA experiments.
PTA experiments function by the longterm monitoring of millisecond pulsars (MSPs), periodically making observations of their highly stable, pulsed radio emission.
One can infer the presence of gravitational waves by looking for correlated deviations in the clock-like behavior of MSPs that are spatially distributed across the galaxy.

By and large, PTAs use traditional pulsar timing methods, which use template-matching techniques to make timestamps on the arrival of pulses from the MSPs, ultimately referencing them to terrestrial clocks and abstract time standards \citep{Hobbs13,Kramer13,Manchester13b,McLaughlin13}.
These fundamental pulsar timing quantities are known as pulse times-of-arrival (TOAs).
Although there have been explorations into methods that effectively bypass the creation of TOAs \citep{Lentati17a}, they still grapple with the issue of template-matching an average pulse profile to data.
Measured TOA uncertainties from PTA MSPs are typically between tens of nanoseconds to several microseconds; the root-mean-square of the residuals from a timing model fit to TOAs collected over many years is currently at the level of a few hundred nanoseconds for many MSPs \citep{Manchester13,Shannon15,Desvignes16,Arzoumanian18b}.
These figures of merit are rough indicators of a PTA's sensitivity to gravitational waves.

However, PTA sensitivity to gravitational waves is a function of both controllable and uncontrollable parameters.
In the latter category are immutable things pertaining to the pulsars themselves (e.g., timing noise \citep{ShanCord10}, ``jitter'' \citep{Oslowski11,Shannon14,Lam16}) and to that which lies between the telescope and pulsar (e.g., the ionized interstellar medium (ISM) \citep{Keith13,Lee14,Levin16,Jones17}, the solar wind \citep{Madison18,Tiburzi18}.  In the former category, there are at least four ongoing approaches to improve sensitivity to gravitational waves.
The first is to increase the cadence of observations, although this has a more substantial effect on the sensitivity to continuous wave sources and not so much on the ability to detect a stochastic background \citep{Ellis12,Arzoumanian14}.
NANOGrav currently employs a mixed observational strategy such that it targets several of its best timed pulsars more frequently than the others expressedly for this purpose \citep{Arzoumanian18b}.
The second approach is to increase the number of pulsars, which is of primary interest for detecting and characterizing the stochastic background of gravitational waves and its anisotropies \citep{Mingarelli13,Siemens13,Chamberlin15,Vigeland16}.
The third attempts to optimize the allocation of resources by choosing time and frequency coverage tailored to individual pulsars' characteristics \citep{Lee12,Lam18b,Lam18c}.
The final strategy is to exploit the various components of the radiometer equation, which dictate how precisely one can measure the TOAs.

Roughly speaking, the contribution to the TOA uncertainty from radiometer noise\footnote{For a recent, comprehensive review of contributions to the total measurement uncertainty in pulsar timing experiments, see \citet{Verbiest18}.}, $\sigma_{\textrm{rad}}$, scales with the system temperature $T_{\textrm{sys}}$, telescope effective area $A_{\textrm{eff}}$, integration length $t_{\textrm{obs}}$, and bandwidth $\Delta f$ as
\begin{equation}
\label{eqn:sigma_rad}
    \sigma_{\textrm{rad}} \propto \frac{T_{\textrm{sys}}}{A_{\textrm{eff}}\sqrt{t_{\textrm{obs}} \Delta f}}.
\end{equation}
Efforts to increase the effective area amount to building new telescopes with substantial collecting area, and in this realm the PTA community anticipates contributions from the recently commissioned Five Hundred Meter Aperture Spherical Telescope (FAST) \citep{Hobbs14}, the ongoing Large European Array for Pulsars (LEAP) project \citep{Bassa16}, and two nascent telescopes, MeerKat \citep{Bailes18} and the Canadian Hydrogen Intensity Mapping Experiement (CHIME) \citep{Ng17}.
However, this approach is ultimately limited by the presence of ``jitter noise'' inherent in pulsar emission \citep{Oslowski11,Shannon14}.
Increasing the integration time can overcome this jitter limit and increase the TOA precision, but this strategy will forever be in competition with logistical constraints of adding more pulsars to observational campaigns that compete for finite telescope resources.
Changing the last two parameters involves substantial receiver development, and progress along this avenue has provided the impetus for our present work.
While reductions in the overall system temperatures are thought to be limited, increasing the fractional bandwidth of receivers in the gigahertz regime has been an area of recent success.

In particular, the Ultra-Broadband Receiver (UBB) at the Effelsberg 100-m Radio Telescope \citep{Beacon}, the Ultra-Wideband Receiver (UWL) at the Parkes 64-m Radio Telescope \citep{Manchester15}, and the planned upgrade to the MeerKat receivers \citep{Kramer15}, are the first truly broadband receivers designed with pulsar timing goals in mind.
The ongoing development, installation, and commissioning of these receivers is paralleled by the development of data acquisition systems that can handle their instantaneous bandwidth \citep{Comoretto12,Prestage15}, as well as research into the question of what is the optimal frequency range to observe an individual pulsar with given characteristics \citep{Lam18b}.

\subsection{The Large Bandwidth Problem}
\label{subsec:large-bw}

The deployment of wideband/broadband receivers for regular PTA observations has necessitated new developments in our timing measurements and analyses, with the reason being twofold.
First, the conventional TOA measurement assumes that the shape of the pulse profile does not evolve with frequency.
A typical protocol is to frequency-average (``f-scrunch'') the profile data and use a template-matching algorithm to measure the band-averaged TOA using a one-dimensional template profile \citep{Manchester13,Desvignes16}.
Another typical protocol is to make ``subbanded'' TOAs by maintaining some of the frequency resolution of the profile data, and measuring a TOA in each subband or channel using the same, constant one-dimensional template profile \citep{Demorest13,Arzoumanian15b,Arzoumanian18b}.
However, it is has been known since the beginning of pulsar observations that pulse profile shapes evolve with frequency due to intrinsic changes in the pulsar magnetosphere as well as extrinsic impressions from propagation through the ISM \citep{Craft68,Craft70}.
Ignoring profile evolution in these ways results in suboptimal template-matching, biased TOAs and timing results, and a loss of precision and sensitivity \citep{Oslowski11,Liu11,Liu14,PDR14,PennucciPhDT,Lentati17b,Arzoumanian15b}.

Alongside of this problem is a more practical one.
As PTAs observe more sources with more telescopes, higher cadence, and broader bandwidths, the volume of TOAs generated is quickly becoming too large for efficient searches for gravitational waves using the current comprehensive Bayesian analyses that, for example, NANOGrav employs \citep{Arzoumanian18a}\footnote{The covariance matrices of PTA datasets are already sufficiently large that low-rank approximations are necessary to make the analyses practical \citep{vH15}.}.
In particular, the need to measure the time-varying dispersion measure (DM) for each pulsar means that large bandwidth observations must be subbanded into smaller frequency chunks, as mentioned earlier, with a TOA produced from each subband.
These are the ``channelized'' or ``subbanded'' TOAs that have comprised the conventional NANOGrav datasets heretofore \citep{Arzoumanian18b}.
For NANOGrav, the typical TOA is measured in a subband with a bandwidth of order $\sim$10~MHz, and thus a single integration from a single pulsar's observation produces dozens of TOAs.
In future observations using upcoming ultra-wide bandwidth receivers, one could expect one hundred or more TOAs per observation, per pulsar, if similar channel bandwidths are used.
Further to this point, should CHIME provide high cadence timing observations for NANOGrav in the near future, we expect to double our already near-unmanageable TOA volume in less than one year.
Therefore, we sought an efficient way to reduce the data volume while optimally using all information in the profile data.

\citet{Liu14} and \citet{PDR14} presented similar methods for solving this ``large bandwidth problem'' \citep{Lommen13}.
The solution involves using a two-dimensional template (pulse amplitude as a function of rotational phase and frequency) that incorporates profile evolution to estimate both the DM within the band in question and a single ``wideband TOA'' at a reference frequency towards the middle of the band. 
In effect, this is similar to estimating the slope and intercept of a line, respectively.
Although both sets of authors explored the tradition of using analytical templates based on Gaussian components whose parameters independently evolve, neither makes a definitive prescription for template generation when using their methods for high precision pulsar timing.
\\\\
\subsection{Beyond Gaussian Templates}
\label{subsec:beyond-gauss}

Decomposing an arbitrary pulse profile shape into a small finite set of analytic basis functions is difficult.
The common approach of using a sum of Gaussian components -- whose parameters can evolve with frequency -- to model a profile has a relatively long history \citep{Krishnamohan83,Kramer94,Lommen01,Liu14,PDR14,PennucciPhDT}.
This approach is limited in applicability and it fails to simply describe the richness of complexity in the evolving profiles of millisecond pulsars.
It is often found in analyses that attempt to infer rough widths or shapes in order to model magnetospheric processes \citep[e.g.,][]{Kramer94b}.
A sum of Gaussian components is also commonly used when estimating scattering parameters, either through forward modeling or backward deconvolution of the pulse shape with the pulse broadening function of the inhomogeneous ISM \citep{Bhat03,Lohmer04,Lewandowski15a,Geyer16,Geyer17}.

However, for purposes of high precision pulsar timing, a high fidelity pulse shape and its evolution with frequency are required, and less important is a physical underpinning that predicts e.g., component shapes and certain functional forms of their evolution.
Small deviations from the true pulse shape lead to inaccurate TOAs, as was already mentioned.
These are often constant biases that can be later modeled and ``corrected'' (e.g., the ``FD'' parameters used in \citet{Arzoumanian15b}, or the ``JUMP'' parameters used in \citet{Demorest13}; see Section~\ref{subsec:timing_dev} herein), but the mismatch of the model to the data also leads to suboptimal TOA uncertainties from the template-matching procedure.

Millisecond pulsars generally have larger duty cycles and more complex profiles than canonical, slow pulsars \citep{Stairs99,Yan11,Dai15,Gentile18}.
Ten or more functional components are often needed to model all of the features of a non-Gaussian shaped profile, even if the profile consists of a single main pulse and interpulse (Table~4.3 of \citet{PennucciPhDT} and plots therein; also see \citet{Lam18e}).
In general, the mapping of analytic functional components to profile components is subjective and uninformative, which is the same as stating that the many parameters from all of the components are highly covariant; such ambiguity undermines physical interpretation.

Finally, a smooth, noise-free template is necessary to avoid subtle biases in the TOAs \citep{Liu11}.
Using a noisy template that was generated from the same data from which the TOAs are measured is sometimes called ``self-standarding'' \citep{Hotan05}.
The primary issue with self-standarding is that, because the noise level in the template does not vanish completely and originates from the data, some of the high frequency noise peaks in the template will be correlated with the noise peaks in the data, thereby adding spurious significance to the cross-correlation.
Ultimately, self-standarding may underestimate the TOA uncertainty and bias its value, particularly in cases of profiles with low signal-to-noise ratios (S/N's) and wide components that will have their TOAs measured using Fourier-domain techniques.

A first attempt at making a two-dimensional template in an analogous fashion to the simple template profiles that are commonly used would involve a similar averaging of all the data, but keeping the frequency resolution, and then smoothing the result.
As we will see, this is unnecessary due to the correlations inherent in profile evolution, and inelegant due to the fact that it does not have a built-in interpolation mechanism for predicting a template profile at any input frequency.

More recently, several authors have investigated methods for template and TOA generation that rely on statistical inference of single pulses, which can help reduce the influence of jitter noise \citep{Imgrund15,Kerr15}.
These more complex methods deviate substantially from classical ones, but do not yet directly address profile evolution with frequency, the issues arising from it, and have not been applied to frequency-resolved, broadband data.
Moreover, they require the detection of individual pulses, which are almost never seen in the MSP population.
\citet{Kerr15} does propose a method for pulsars with undetected single pulses, the application of which would be interesting to see on broadband PTA datasets.

It is apparent, then, that it is difficult to design a robust protocol that returns a model of pulse profile evolution which is both simple and accurate. 

In this paper we propose a new approach to produce high fidelity, noise-free template profiles at any frequency within an observed bandwidth for the primary objective of precise TOA generation\footnote{This new method is used in the generation of NANOGrav's forthcoming 12.5-year  ``wideband'' dataset, which will be presented elsewhere.}.
The basic assumption in our method is that profile evolution is smoothly and slowly varying with frequency; this long-standing observation was stated in similar words at least as early as \citet{Komesaroff70}.
Put another way, a profile observed at frequency $\nu$ is arbitrarily similar to a profile observed at $\nu+\delta\nu$ for arbitrarily small $\delta\nu$, and profiles at nearby frequencies are highly correlated in shape.
This assumption is validated empirically, and it reflects the theoretical picture that the radio emission at different frequencies corresponds to different radial depths in the magnetosphere, in which the emitting plasma accelerates over a smoothly varying magnetic field \citep[][and references therein]{Chen14}.

In the next section (Section~\ref{sec:modeling}), we describe our new protocol for modeling profile evolution and making noise-free phase-frequency templates. We have chosen the MSP \epsr\ as a demonstration pulsar because it is included in all of the current PTA experiments, it is one of the best timed pulsars (due to the combination of its brightness, duty cycle, spin period, and stability), it scintillates strongly around the commonly observed frequency 1.5~GHz, and because it has highly significant, albeit subtle, profile evolution.
Because of the relatively large fractional bandwidth ($\sim$0.5), we will model \epsr's profile evolution across the $\sim$800~MHz bandwidth centered within L-band (1.5~GHz) as seen by the GUPPI backend instrument \citep{Duplain08} at the 100-m Robert C. Byrd Green Bank Telescope (GBT).
The details of the GBT data used can be found in \citet{Arzoumanian15b,Arzoumanian18b} and in the forthcoming NANOGrav 12.5-year data release paper.
We conclude with a discussion and future prospects in Section~\ref{sec:conclusion}.

\section{Modeling Profile Evolution}
\label{sec:modeling}
\makeatletter{}In most observations of known pulsars, the observed bandwidth from the lowest frequency $\nulo$ to the highest frequency $\nuhi$ is broken up into $\nchan$ frequency channels, and the time-series data are folded modulo a predetermined timing model into $\nbin$ pulse profile phase bins.
A single phase-frequency observation is often referred to as a ``subintegration'', but we will also use the word ``portrait'' to refer to any set of continuously changing pulse profiles (e.g., Figure~\ref{fig:examp-port}).

Assume one has a noise-free observation of the average portrait (i.e., the collection of average profiles obtained simultaneously from a single broadband instrument) with arbitrarily good phase and frequency resolution, from a pulsar with a DM of zero, and further assume the average portrait does not vary from observation to observation.
We imagine each of the $\nchan$ profiles as a vector with $\nbin$ coordinates.
Because profile evolution is smoothly varying, the tips of these vectors trace out some curve; in a sense, this arbitrary geometric curve \textit{is} the profile evolution for a particular pulsar.
We seek to parameterize this curve by frequency, extending from $\nulo$ to $\nuhi$, so that we can evaluate it at any desired value on the interval and obtain a noise-free template for measuring TOAs.
This nominally entails characterizing all $\nbin$ coordinate functions.
However, the vectors only span a maximum of $\min(\nchan,\nbin)$ dimensions, and typically $\nchan < \nbin$\footnote{A collection of $n_{\textrm{\scriptsize{chan}}}$ vectors span $n_{\textrm{\scriptsize{chan}}}$ dimensions if and only if they are linearly independent, otherwise they span fewer than $n_{\textrm{\scriptsize{chan}}}$ dimensions.  If the vectors are represented by $n_{\textrm{\scriptsize{bin}}}$ coordinates and $n_{\textrm{\scriptsize{bin}}} < n_{\textrm{\scriptsize{chan}}}$, they cannot be linearly independent (e.g., consider any three non-zero vectors in the usual Euclidean plane), and thus span a maximum of $n_{\textrm{\scriptsize{bin}}}$ dimensions.  The method will work in either case.}.
Moreover, if pulse profiles evolve slowly, then we should be able to find a set of $\neig << \nchan$ basis vectors that span the majority of systematic variation among the profiles.
We therefore decompose the average portrait into an orthonormal set of eigenvectors, or what we will call ``eigenprofiles'', using Principal Component Analysis (PCA).
A model for profile evolution is created by selecting a small number of eigenprofiles to use as a basis for the aforementioned curve, and a spline is fit to each coordinate function corresponding to the profiles projected onto the denoised basis profiles.

The remainder of this section explains the process in detail.

\subsection{Assembling an Average Portrait}
\label{subsec:avg-port}

In a conventional analysis, an average profile is made by incoherently shifting and summing all significantly detected total intensity profiles from all frequencies in a particular band.
This process is iterative, often starting with some fiducial or arbitrary pulse shape against which all profiles are aligned using a typical phase-gradient algorithm \citep[cf. Section~2.4 of][]{DemorestPhDT}.
The result is then used as a starting reference profile, the process is repeated several times, and the final result is smoothed/denoised and used as a template for measuring TOAs.

In making an average phase-frequency portrait, we execute these same steps and use the final result as a starting point for our alignment procedure.
That is, we initially assume zero profile evolution by aligning each phase-frequency subintegration in the dataset (a ``data portrait'') relative to a constant-profile portrait comprised of the final, smoothed average profile.
Instead of aligning each data portrait relative to the constant-profile portrait using only an overall, achromatic phase offset, each profile is additionally rotated by an amount proportional to the inverse-square of its frequency.
This is tantamount to the fitting procedure described in \citet{PDR14} used to simultaneously measure a TOA and DM per subintegration. In this way, we attempt to remove the dispersive delays from DM variations that could result in smearing of the average portrait.

For example, the DMs of pulsars can change by several $\times\ 10^{-4}-10^{-3}$~cm$^{-3}$~pc over many years \citep{Jones17}; for our example data and pulsar, this is especially true \citep{Lam18}, and could induce rotational differences between the top and bottom of the band by up to one phase bin\footnote{NANOGrav collects profiles with 2048~phase bins, so for a spin period of $\sim$4~ms, one phase bin corresponds to $\sim$2~$\mu$s.}.
Although this is a small difference, the uncertainties on the DM measurements are often $<$1$\times 10^{-4}$~cm$^{-3}$~pc (corresponding to a fraction of a phase bin across the band), and so the differences between DMs can often be detected.
In any event, making note of the practice is warranted in the case of datasets with larger fractional bandwidth.

The aligned data portraits are thus averaged together, the process is similarly iterated over, and a final, noisy average portrait is obtained.
Note, this process does not disentangle profile evolution neither from the absolute DM nor from phase delays that scale differently than $\nu^{-2}$, and so, in a strict sense, the alignment will be wrong; to first order, it will be wrong to within a small rotation that scales as $\nu^{-2}$ (also see Section~\ref{subsec:future}).

Because we want the model to capture pulse shape changes and not overall intensity variations from the pulsar's spectral index or diffractive scintillation that has not been averaged out, we normalize each individual profile of the average portrait as follows.
Each channel's profile is scaled by an individual maximum-likelihood amplitude parameter from a matched-template fit using a phase-gradient algorithm, using the mean profile as the template.
See, for example, Equation~2.3 of \citet{DemorestPhDT} for how to obtain this normalization factor.
No phase shifts are applied in this normalization and note that the mean profile, which will be used in the next section, will have changed depending on the choice of normalization.
Other normalizations are of course possible, and the main effect from not normalizing at all is that there will be an additional eigenvector (see Section~\ref{subsec:pca}) that may strongly resemble the mean profile.
Recall, when measuring a ``wideband TOA'' from an individual observation, the specific intensity variations of that subintegration will be captured by the model via the individual channel amplitude parameters $\an$ \citep[Equation~11 of][]{PDR14}, and thus the particular choice of normalization is mostly inconsequential.

However, care should be taken in cases when strong diffractive scintillation is still evident in the average portrait such that parts of the band have very little or no signal.
Using an alternate normalization than the one mentioned or manually zero-weighting (``zapping'') the average portrait where there is no signal ensures that the procedure will be uncorrupted.

\begin{figure}
\begin{center}
\includegraphics[width=\columnwidth]{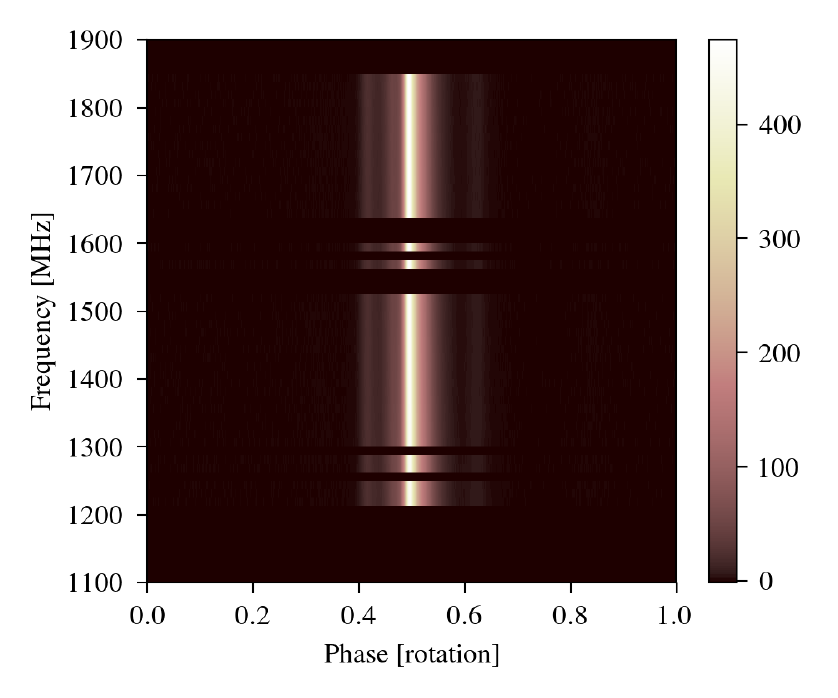}
\caption{
Example average portrait for \epsr\ observed at L-band with the GBT using GUPPI (1.5~GHz with 800~MHz bandwidth).
Gaps represent zero-weighted channels that were consistently or strongly contaminated by RFI.
Each channel has been normalized by the mean profile (Section~\ref{subsec:avg-port}); for this reason, the amplitude scale, here and elsewhere, is unitless.
By eye, there is little or no obvious profile evolution, and one might surmise that using a single, average profile (e.g., the top panel of Figure~\ref{fig:examp-eigvec}) would be sufficient for timing.
However, the profile evolution is prominently displayed in the difference portrait of Figure~\ref{fig:examp-diff}.
}
\label{fig:examp-port}
\end{center}
\end{figure}

We carried out the above procedure on the fully processed and calibrated GUPPI L-band (1.5~GHz) observations of \epsr\ contained in NANOGrav's 12.5-year dataset (to be published elsewhere).
The average portrait is plotted in Figure~\ref{fig:examp-port}, which has $\nbin = 2048$ and, of the 64 original channels, there are $\nchan = 42$ channels that have not been zero-weighted.  Conservative zapping does not negatively affect the modelling (since, as we will see, $\neig << \nchan$), while it best removes contamination, e.g., radio frequency interference (RFI), that could show up as a spurious eigenvector (see Section~\ref{subsec:pca}).
The individual average profile S/N's lie between $\sim$5500--10000, with a median of $\sim$7800.
Jitter noise will be present in any such average portrait and, in principle, can bias the following modeling procedure.
However, Figure~\ref{fig:examp-port} represents an average of approximately 250 integrations of 20--30~minutes each.
The 30-min integration jitter noise for \epsr\ at L-band in our dataset has been measured to be tens of nanoseconds \citep{Lam18e}.
Therefore, the corresponding amount of jitter present in this average portrait is no more than a few nanoseconds, corresponding to less than 1\% of a phase bin, and so we expect the following results to be unaffected by jitter.

\makeatletter{}

\subsection{Principal Component Analysis\\of the Average Portrait}
\label{subsec:pca}

PCA has been used a handful of times in pulsar timing analyses before, but only in the capacity of characterizing \textit{temporal} profile variability \citep{BlaskiewiczPhDT,Oslowski11,DemorestPhDT,Stairs04,Lin18}.
As stated, we assume that the average portrait does not vary in time.
This is a reasonable assumption because it has not been shown that temporal variability of intrinsic average profile shapes is a limiting factor in the timing of millisecond pulsars \citep{Shao13,Shannon16,Brook18}.
At the same time, this assumption highlights an important drawback of our modeling technique, as it makes no attempt to disentangle profile shape changes intrinsic to the pulsar and pulse broadening from the interstellar medium.
For example, such a model could not be easily used if one is attempting to measure time-varying scattering effects \citep[][Pennucci et al. (in prep.)]{Lentati17b}.
However, see Section~\ref{subsec:future} for more discussion on these matters.

\begin{figure}
\begin{center}
\includegraphics[width=\columnwidth]{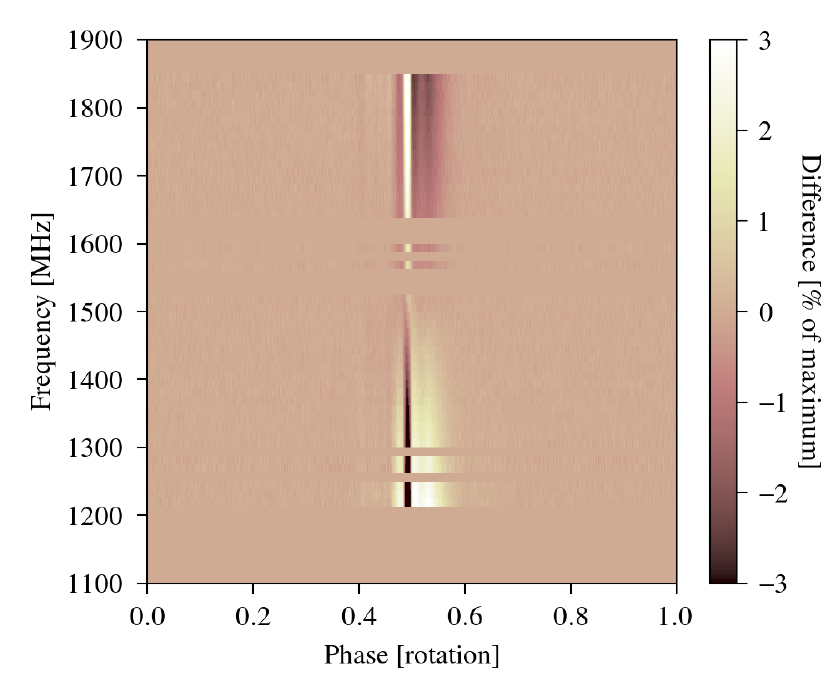}
\caption{
The difference portrait $\mathbf{\Delta P}$, which is the difference between the data from Figure~\ref{fig:examp-port} and its weighted mean profile, which is shown in the top panel of Figure~\ref{fig:examp-eigvec}.
A very similar result is obtained if the mean profile is instead template-matched to each channel's profile before the difference is calculated.
Profile evolution is evident, which underscores the shortcoming of template-matching using a mean profile.
Note that the difference amplitude has been scaled here to percent of the maximum value in Figure~\ref{fig:examp-port}.
}
\label{fig:examp-diff}
\end{center}
\end{figure}

We follow a standard procedure to carry out the PCA.
For the $\nchan \times \nbin$ portrait of row-vector profiles $\mathbf{P} = \{p_1, p_2, ..., p_{n_{\textrm{\scriptsize{chan}}}}\} = \{p_n\}$, with weighted mean profile given by
\begin{equation}
\label{eqn:pn}
    \overline{p} = \big{(} \sum_n \sigma^{-2}_n p_n\big{)} \big{(} \sum_n \sigma^{-2}_n \big{)}^{-1}, 
\end{equation}
the weighted covariance matrix $\mathbf{C}$ is formed as
\begin{equation}
\label{eqn:Cmm}
    \mathbf{C} = \big{(} \sum_n ( \sigma^{-2}_n \mathbf{\Delta P}^\transpose \mathbf{\Delta P} ) \big{)} \big{(}\sum_n \sigma^{-2}_n\big{)}^{-1},
\end{equation}
where $\mathbf{\Delta P}$ is the difference portrait $\{p_n - \overline{p}\}$ (e.g., see Figure~\ref{fig:examp-diff}), and $\sigma_n$ is the noise level in channel $n$, which we estimate from the last quarter of the profile's power spectrum.
$\mathbf{C}$ is diagonalized numerically such that the orthogonal transformation $\mathbf{O}^{-1} = \mathbf{O}^\transpose$ has basis row-vectors that are our ``eigenprofiles'', and by convention they are sorted according to the magnitude of their corresponding eigenvalue, $\lambda_m$:
\begin{equation}
\label{eqn:lambdam}
    \textrm{Diag}(\lambda_m) = \mathbf{O}^{-1} \mathbf{C} \mathbf{O}.
\end{equation}
The magnitude of each eigenvalue corresponds to the amount of variance among the profiles projected along the corresponding eigenprofile basis vector.
However, we will find that using the eigenvalues alone to select significant eigenvectors will prove insufficient.

PCA is often employed to explore statistical variance within a sample that might randomly occupy an elliptical volume in the vector space of principal components. Here we are interested in using it, ideally, to encapsulate only the systematic variance arising from the slow evolution of highly correlated samples, but in practice all of the data, including the average portrait, are contaminated by (at least) radiometer noise.
Assuming the common situation where $\nchan < \nbin$, this added randomness to the vector components implies strict linear independence among the profiles, even in the absence of any profile evolution\footnote{To see why this is true, consider the simple case of zero profile evolution, in which case all of the vectors are identical (and, thus, linearly dependent).  If you add random numbers to each component of each vector, it is exceedingly unlikely that any three vectors lie in the same plane, and thus no single vector can be expressed as a linear combination of the others.}, in which case the PCA will return $\nchan - 1$ noisy eigenprofiles with non-zero eigenvalues\footnote{The subtraction of the weighted mean profile is responsible for this number being one less than $n_{\textrm{\scriptsize{chan}}}$.}.
In the case $n_{\textrm{\scriptsize{chan}}} > n_{\textrm{\scriptsize{bin}}}$, which is more typically the case in PCA (i.e., the number of ``observations'' exceeds the number of ``variables''), the profiles cannot be linearly independent, as mentioned earlier, but the random component similarly ensures that the PCA will return $n_{\textrm{\scriptsize{bin}}}$ eigenprofiles.
In either case, the added noise presents two problems.
First, any reconstructed profile made from a linear combination of these eigenprofiles will also be noisy, and we are interested in smooth templates.
Second, determining the number of ``significant'' eigenprofiles $\neig$ that sufficiently capture the variance in the data besides that induced by statistical noise, and from which we will construct a model, is a common problem in PCA; of course, we want to keep $\neig$ small.
We attempt to solve both of these issues by following the analogous practice when forming a one-dimensional template profile: we smooth the eigenprofiles as well as the mean profile.
In this way, we ensure that any reconstructed template is smooth, and smoothing the eigenprofiles permits us to estimate their S/N, with which we can judge their significance.

\makeatletter{}

\subsection{Profile Smoothing}
\label{subsec:smoothing}

We use a Stationary Wavelet Transform \citep{pywt} to denoise the mean and eigenprofiles; this type of wavelet transform is independent of a phase offset of the input.
The results will vary based on the choice of wavelet family and thresholding procedure, which we omit to specify without loss of generality in the present discussion.
However, for a given choice, we determine the level of wavelet decomposition and thresholding factor by maximizing the S/N within a fixed tolerance around a reduced chi-squared value of 1.0.
In this way we ensure a balance between under-smoothing, retaining sharp intrinsic features, and over-smoothing.
The S/N metric we use is defined in Appendix~A of \cite{Arzoumanian15b}; the signal is calculated from the power spectrum of the input profile, and the noise level is estimated from the last quarter of the power spectrum.

\subsection{Eigenvector Selection}
\label{subsec:eigvec}

\begin{figure}
\begin{center}
\includegraphics[width=1.0\columnwidth]{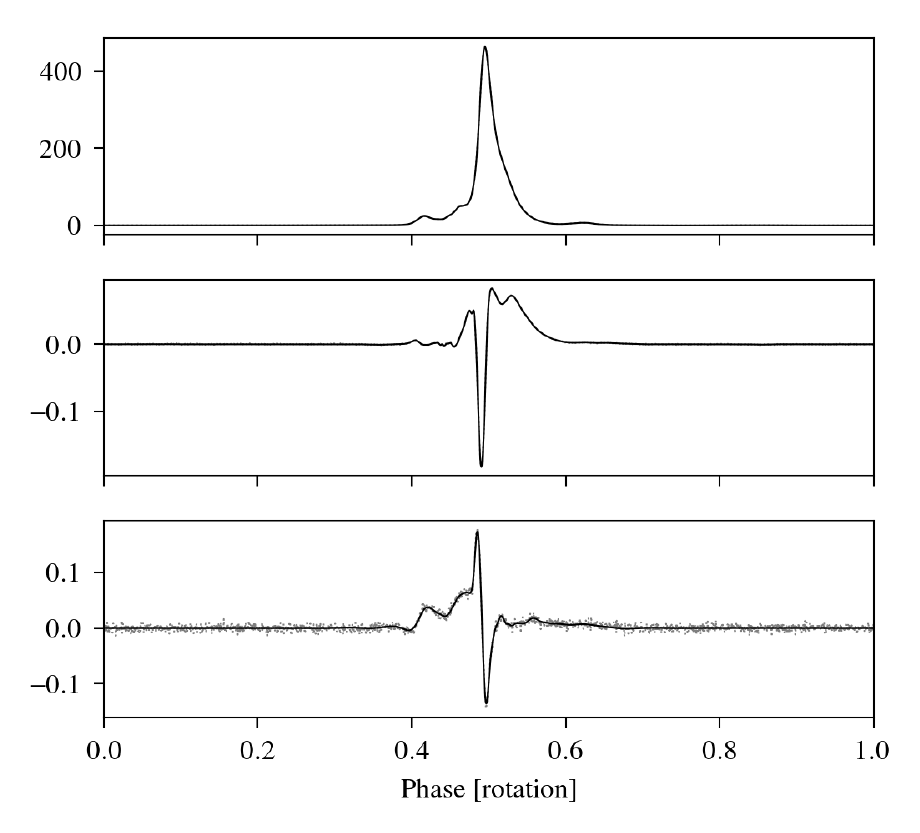}
\caption{
The mean profile (top panel) and eigenprofiles (lower two panels) corresponding to the decomposition of the \epsr\ data from Figures~\ref{fig:examp-port}~\&~\ref{fig:examp-diff}.
The gray points, only discernable in the bottom panel, are the values computed from the data, whereas the black lines are their smoothed counterparts that comprise part of the model.
}
\label{fig:examp-eigvec}
\end{center}
\end{figure}

A simple attempt to determine $\neig$ based on the magnitude of the eigenvalues will, in some cases, miss significant eigenvectors while keeping ``pure-noise'' eigenvectors for the reason mentioned earlier; the variance from radiometer noise (or other contamination) may be larger than the intrinsic, systematic change along an eigenprofile that actually encompasses profile evolution.
The trivial example is that it is difficult to design a robust selection procedure based on the eigenvalues' magnitudes alone that will return $\neig = 0$ when there is no intrinsic profile evolution.
For this reason, we determine $\neig$ by ordering the eigenvectors based on their S/N, and retain up to ten eigenprofiles above a certain threshold\footnote{One expects the eigenvalues and their corresponding eigenvector S/N's to be correlated, but for the reasons given it is better to select based on eigenvector S/N.}.
The S/N metric used is the same as that mentioned in Section~\ref{subsec:smoothing}.
Figure~\ref{fig:examp-eigvec} shows the mean profile, eigenprofiles, and their smoothed counterparts for the data from Figures~\ref{fig:examp-port}~\&~\ref{fig:examp-diff}.
\makeatletter{}

\subsection{Interpolation of the Projected Profile Coordinates}
\label{subsec:interp}

The final step is to project the $\nchan$ mean-subtracted pulse profiles onto the basis of smoothed eigenprofiles and to find a set of $\neig$ interpolating coordinate curves, each parameterized by frequency $\nu$.
The data profiles expressed in this subspace are given by\begin{equation}
\label{eqn:P_prime}
        \mathbf{P}' = \mathbf{\Delta P}\mathbf{\widetilde{O}},
\end{equation}
where $\mathbf{\widetilde{O}} = \{\hat{e}_1, \hat{e}_2, ..., \hat{e}_{\neig}\} = \{\hat{e}_i\}$ is the matrix containing as its columns a selection of $\neig$ basis eigenprofiles from $\mathbf{O}$ that have been wavelet smoothed.
Each column of $\mathbf{P}'$ contains a sequence of coordinates to be parameterized.

We approximate each coordinate curve by fitting a smoothly varying piecewise polynomial spline function, represented as a sum of basis splines (``B-splines'' \citep{Schoenberg46}).
B-splines are so called because for a given a set of unique knot locations (called ``breakpoints''), a set of continuity conditions at the breakpoints, and the polynomial degree of the interpolating spline $k$\footnote{To avoid confusion, the polynomial degree corresponds to the highest exponent in the polynomial, whereas the polynomial order is equal to the number of coefficients that determine the polynomial, here $k+1$.
These terms are sometimes used ambiguously in the literature; we use $k=3$ cubic splines.}, any spline function can be expressed as a unique linear combination of the corresponding B-splines.
For a concise review of B-splines, see Sections 1 \& 2 of \citet{Bachau01} \citep[or, for a comprehensive guide,][]{deBoor78}.
We make the common, simplest choice of maximum continuity on the open interval $(\nulo, \nuhi)$, which means that internal breakpoints (non-endpoints) each have one knot only.
That is, these knots have multiplicity of one, whereas the unimportant endpoint knots will have maximal multiplicity $= k+1$ and no continuity.
For example, for $k = 3$ cubic splines, this means the splines have continuous first and second derivatives at internal breakpoints.

With these choices of continuity and polynomial degree, the B-splines are uniquely determined (up to a multiplicative constant) for a given sequence of breakpoints that divide the bandwidth into $l$ segments.
The number of B-splines $\nB$ is then given by
\begin{equation}
\label{eqn:nB}
    \nB = l + k = \nt - k - 1.
\end{equation}
The number of knots $\nt$ and their locations $\{t_1, t_2, ... t_{\nt}\} \\= \{t_q\}$, in turn, are determined by how smooth the interpolating curve is desired to be.
This part of the procedure is completed numerically by the \texttt{splprep} routine in \texttt{SciPy} \citep{splprep,scipy}.
The routine minimizes a sum-of-squared-deviations metric to meet the desired smoothness criterion, and returns the knot locations $\{t_q\}$ and B-spline coefficients $c_{ij}$.
Again, we leave the amount of smoothness undetermined in our implementation, but it can also be specified in reverse: the maximum number of unique knots allowed can be specified, where fewer knots obviously means more smoothing and, thus, fewer B-splines.

As an example, for $k = 3$ with maximum internal continuity and smoothness, there will be only two breakpoints (the endpoints $\nulo$ and $\nuhi$) and one interval between them, $l = 1$.
This corresponds to a multiplicity of $k+1 = 4$ knots at each end breakpoint.
Therefore, $\nt = 8$, giving the minimum number of B-splines for $k = 3$, which is $n_{\textrm{\scriptsize{B}}} = 4$.

The fit to the coordinate sequences in $\mathbf{P}'$ is performed globally over a single shared set of knots, and returns $\neig$ parametric equations describing a single profile evolution curve.
The parametric equations $S_i$ can be evaluated as
\begin{equation}
\label{eqn:Bspline}
    S_i(\nu) = \sum^{n_{\textrm{\scriptsize{B}}}}_{j=1} c_{ij} B_{tk,j}(\nu).
\end{equation}
The explicit formulae for the piecewise polynomial B-spline functions $B_{tk,j}$ are never required, as they are numerically determined by the software when given as input a set of knots, continuity conditions, and polynomial degree (the subscript $tk$ is to emphasize that the splines depend on the choice of knots and polynomial degree).

\begin{figure}
\begin{center}
\includegraphics[width=\columnwidth]{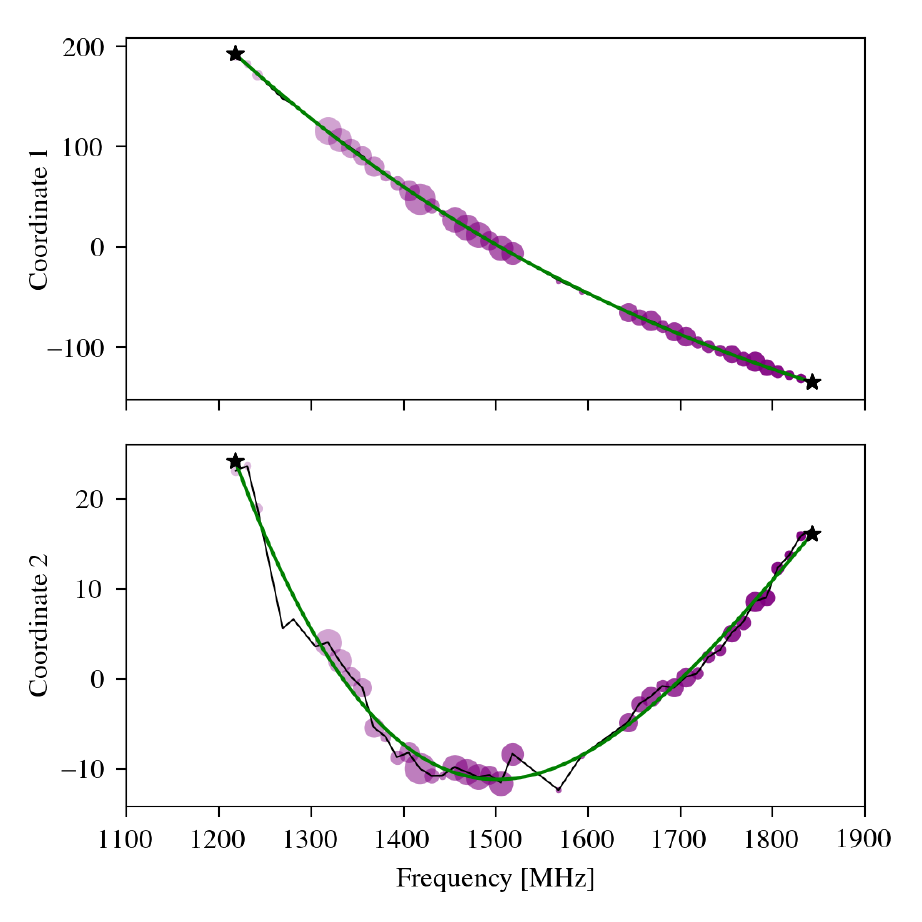}
\caption{
Coordinate curves parameterized by frequency for \epsr.
The purple points correspond to the difference profiles from Figure~\ref{fig:examp-diff} projected onto the eigenprofiles of Figure~\ref{fig:examp-eigvec} (i.e., the columns of $\mathbf{P}'$); the black line simply connects the points for guidance.
The green curve is the spline model of profile evolution, $S_i(\nu)$, interpolating values continuously across frequency.
The size of the points corresponds to the profile's S/N (used as weights in the spline regression), whereas their shading indicates the profile's frequency (see Figure~\ref{fig:examp-proj}).
The small black stars at the ends are the spline break points; templates generated outside of their range may quickly become inaccurate.
}
\label{fig:examp-freq}
\end{center}
\end{figure}

\begin{figure}
\begin{center}
\includegraphics[width=\columnwidth]{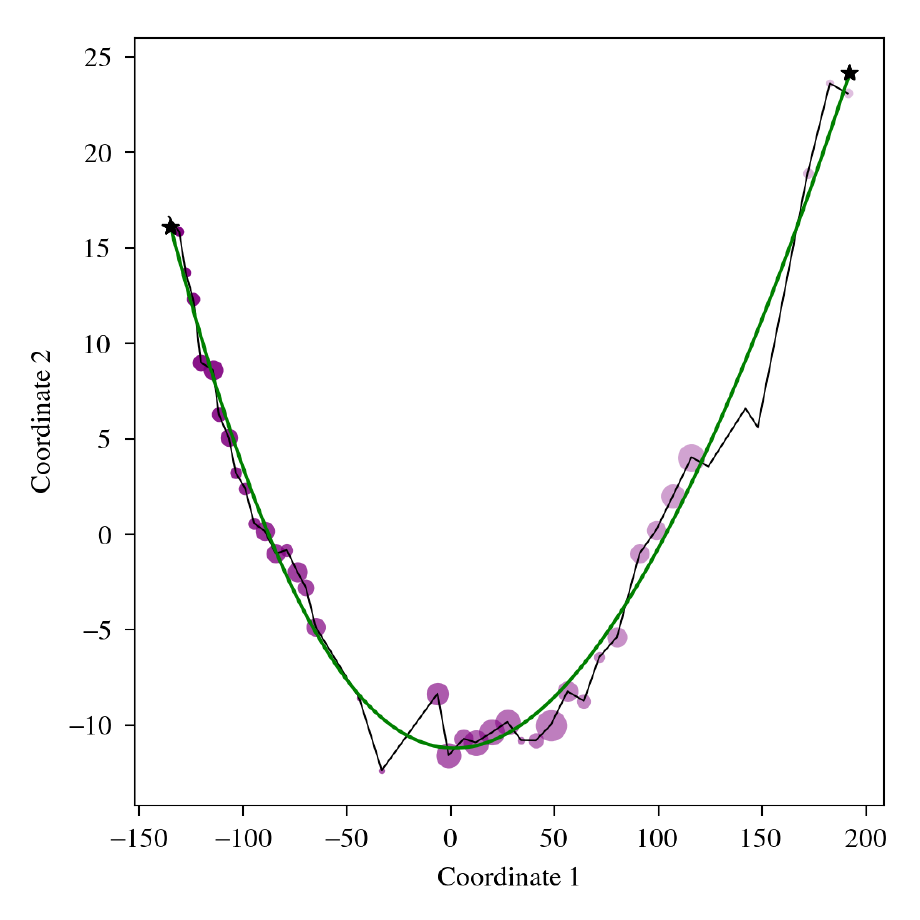}
\caption{
The profile evolution of \epsr\ between $\sim$1200~MHz (lightest shade)  and $\sim$1850~MHz (darkest shade) represented by the frequency-parameterized curves of Figure~\ref{fig:examp-freq}.
The two smoothed eigenprofiles of Figure~\ref{fig:examp-eigvec} form the (approximately) orthonormal basis for the projected subspace plotted here.
The coordinate origin corresponds to the tip of the mean profile vector.
}
\label{fig:examp-proj}
\end{center}
\end{figure}

Figures~\ref{fig:examp-freq}~\&~\ref{fig:examp-proj} show the parametric equations for the profile evolution model of \epsr\ around 1.5~GHz.
These figures help to highlight some advantages of our method over using a simple average portrait as a template, in addition to the argument based on parsimony in the next section.
When simply smoothing individual profiles from an average portrait, each template profile's shape is a function only of information in that one profile instead of being predicted from the correlations inherent in profile evolution, as is done here.
For instance, smoothing an average profile that has a low S/N from strong diffractive scintillation (e.g., one of the profiles corresponding to the smallest purple points in the figures, although all of these profiles have large S/N) often produces unsatisfactory results in the form of artificial peaks or fluctuating baselines.
However, in our method we are able to accurately predict these template profile shapes by leveraging the information contained in the evolution of high S/N profiles; the interpolated coefficients are used to linearly sum the highly significant, smoothed eigenprofiles (see Equation~\ref{eqn:Tprof}), resulting in a noise-free template profile free of artifacts.
In fact, the model will be able to reliably interpolate and predict template profiles in gaps of the data, such as those arising from RFI (e.g., around 1550~MHz in our example) or those between scintillation maxima.
\makeatletter{}
 
\makeatletter{}

\subsection{Model Summary \& Template Construction}
\label{subsec:model-const}

In summary, a model of profile evolution is composed of a denoised mean profile for the frequency band in question $\widetilde{p}$ ($\nbin$ numbers), a matrix of denoised eigenprofiles $\mathbf{\widetilde{O}}$ ($\nbin \times \neig$ numbers), a set of increasing knot locations in frequency $\{t_q\}$ ($\nt$ numbers), the matrix of B-spline coefficients $c_{ij}$ ($\neig \times \nB$ numbers), and the polynomial degree $k$ (one number).
Generally, we have found that $l \sim 1-$few, $\nB \sim 4-$few, $\neig \sim 0-$few, and typically $\nbin \sim$ hundreds--thousands.
Therefore,
\begin{equation}
\label{eqn:nnum}
    (\nbin\times\neig) + (\neig\times\nB) \ + \ \nt \ + \ 1 \approx \nbin \times \neig, \end{equation}
and so there are approximately $\neig$ times more parameters in this description of profile evolution, compared to the conventional model of using a single smoothed mean profile ($\nbin$ numbers) as a constant template.
This makes sense intuitively, and considering we are describing $\nchan \times \nbin$ data points, and $\neig << \nchan$, this is a modest increase.
Consequently, our modeling technique should be preferred over using an $\nchan \times \nbin$ average data portrait, smoothed or not, based on parsimony (``Occam's Razor'').
If this is not immediately clear, consider that the number of eigenprofiles can be incrementally increased and that the spline smoothing parameter can be tuned until the template matches the input data to a vanishingly small difference.

A model takes as input a vector of frequencies $\nu_n$ and a number of desired phase bins.
It reconstructs a smooth template portrait $\mathbf{T}$, which is by obtained by evaluating the coordinate curves at the input frequencies $\mathbf{S} = \{S_i(\nu_n)\}$, calculating the deprojected difference profiles,\begin{equation}
\label{eqn:deltaT}
    \mathbf{\Delta T} = \mathbf{S}^\transpose \mathbf{\widetilde{O}}^\transpose,
\end{equation}
and adding the smoothed mean profile $\widetilde{p}$ to each.
More simply, a single template profile $T$ at frequency $\nu$ is constructed as
\begin{equation}
\label{eqn:Tprof}
    T(\nu) = \sum^{n_{\textrm{\scriptsize{eig}}}}_{i=1} \sum^{n_{\textrm{\scriptsize{B}}}}_{j=1} c_{ij} B_{tk,j}(\nu) \, \hat{e}_i ~ + ~ \widetilde{p}.
\end{equation}
A resampling function is used in case the desired number of phase bins is different from the original $\nbin$.

\begin{figure}
\begin{center}
\includegraphics[width=1.0\columnwidth]{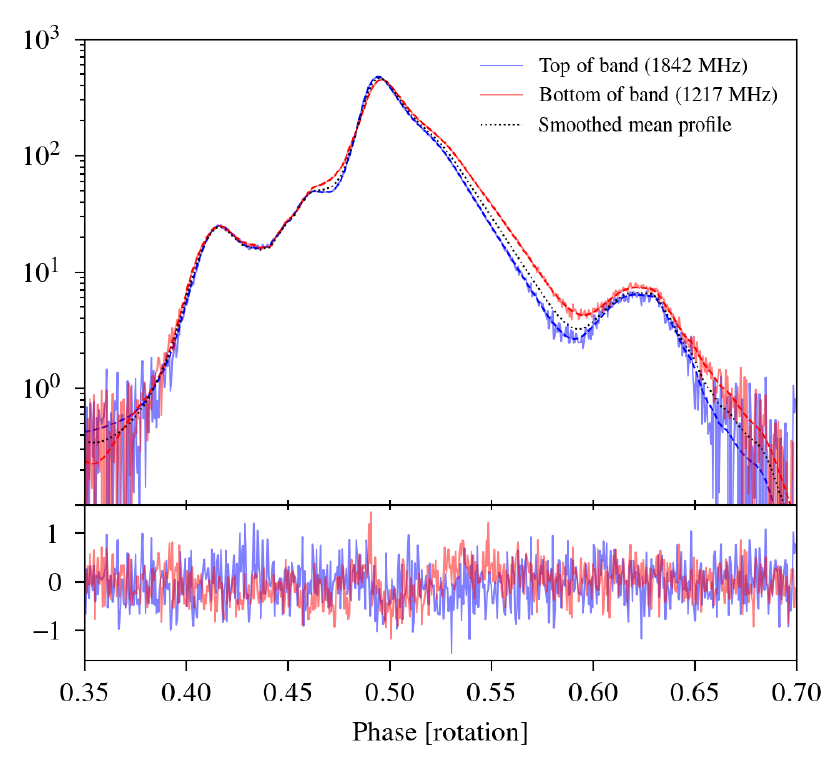}
\caption{
\epsr\ template profiles (dashed lines) overlaid on the average data profiles (light solid lines) from the top (blue) and bottom (red) of the band plotted in Figure~\ref{fig:examp-port}; the bottom panel shows the residuals.
The amplitude is shown on a logarithmic scale, and only the central 35\% of phase -- the on-pulse region -- is displayed.
The minute differences in shape amount to a difference of $\sim$40~$\mu$s across the band when using a smooth mean profile (black dotted line) as a template to measure subbanded TOAs \citep[see Figure~2 of][]{Arzoumanian15b}.
}
\label{fig:examp-model}
\end{center}
\end{figure}

Figure~\ref{fig:examp-model} shows an example of two such reconstructed template profiles, at the extrema of the band from Figure~\ref{fig:examp-port}, along with the original average profiles and the smoothed, band-averaged profile.
The evolution of the profiles across this 600~MHz bandwidth (having a fractional bandwidth of $\sim$0.4) is subtle, but highly significant.
The mean profile does not deviate tremendously in shape from the example profiles, but when it is used to measure subbanded TOAs across such a bandwidth, there is approximately $\sim$40~$\mu$s of constant timing difference that needs to be accounted for in the timing model with additional parameters \citep{Arzoumanian15b}.
The need for such parameters goes away by directly modeling profile evolution, as is the case here.
Besides affecting the accuracy of TOA measurements, because these templates actually match the data, the uncertainties from either subbanded or wideband TOA measurements will be more reliable and, in some cases, improved \citep{PDR14}.
It is expected that these improvements have a positive effect on the precision timing of pulsars, but quantifying these improvements across a range of template generation techniques, timing and noise model choices, and pulsar variety, goes beyond the scope of this paper (however, see the discussion in Section~\ref{subsec:timing_dev}).
\makeatletter{}

Even though in our explorations thus far we have mostly found $\neig = 1-2$ and $l = 1$ (see the forthcoming NANOGrav 12.5-year data release), the method is general and robust enough for much larger fractional bandwidths, which will encompass greater profile evolution.
Thus, with additional testing on real or simulated data we may find models can have several eigenprofiles and perhaps more complex evolution in the space spanned by them.
\\\\

\section{Discussion}
\label{sec:conclusion}
\makeatletter{}In this paper, we have described a generalization of the conventional method for making noise-free profile templates for the purpose of high precision pulsar timing in the era of ultra-wideband receivers.
This is necessitated by the combination of effects on pulsar timing that arise when an incorrect model of pulse profile evolution is used over large bandwidths in the measurement of TOAs.
For instance, unmodeled profile evolution compounded with diffractive interstellar scintillation can stochastically bias subbanded or band-averaged TOAs, since the portions of the observed frequency band that get weakened or amplified (the ``scintles'') vary stochastically \citep{Liu11,Oslowski11}.

\subsection{Synopsis}
\label{subsec:synopsis}

Our new method is a natural extension of methods already being used for generating one-dimensional profile templates in that it follows analogous procedures of averaging and smoothing, and also ultimately produces a smooth mean profile template; it adds only a few smooth eigenvector profiles whose linear combination varies with frequency based on some prescription.
The general protocol, with some of our particular choices proposed in this paper, is outlined as:
\begin{enumerate}[label=\textbf{\arabic*.}]
    \setlength\itemsep{0.25em}
    \item{\textbf{Average the phase-frequency data semi-coherently and iteratively --}
          This step is analogous to what is already typical, but here it is necessary to maintain frequency resolution instead of averaging each subintegration across the bandwidth.
          ``Semi-coherently'' means that instead of aligning each subintegration using a phase offset only, we also include a frequency-dependent rotation per subintegration that effectively accounts for DM variability.}
    \item{\textbf{Decompose the averaged data using Principal Component Analysis --}
          PCA allows us to find a reduced basis in which the majority of the profile evolution occurs.
          Some normalization procedure should first be carried out on the average portrait, and we choose a normalization based on the mean profile.}
    \item{\textbf{Select significant eigenvector profiles --}
          This step will indicate how many orthogonal elements (in addition to the mean profile) will be needed to capture the profile evolution.
          Many possible selection algorithms are possible; we select based on a simple S/N metric.}
    \item{\textbf{Smooth the mean \& eigenvector profiles --}
          Proper smoothing ensures that noise-free, high fidelity templates will be constructed from the mean and eigenvector profiles.
          We choose wavelet smoothing, which is already conventional.}
    \item{\textbf{Choose interpolating coordinate functions --}
          This step prescribes the frequency evolution of the projected coordinate coefficients used in the linear combinination of the eigenprofiles.
          Again, there are many possibilities, but we find that B-splines are convenient for this purpose.}
\end{enumerate}
Thus, in general, a model consists of smoothed mean and eigenvector profiles, plus a set of coefficients or other parameters that describe how to analytically generate the coordinates for linearly combining the eigenvector profiles given an input frequency.

\subsection{Wideband Timing}
\label{subsec:timing_dev}

In principle, our templates could be used in various ways for conventional TOA measurements and timing analyses.
For instance, the individual template profiles could be used to make subbanded TOAs.
If desired, the model alignment could then be arbitrarily altered either by using numerous TOA phase offsets (``JUMPs'') as free parameters in the timing model \citep[cf.][]{Demorest13}, or by using some functional relationship to introduce fixed (or variable) phase delays as a function of frequency \citep[e.g., ``FD'' parameters; cf.][]{Arzoumanian15b,Arzoumanian18b}.
However, the suggested use of these models is for the measurement of wideband TOAs \citep{Liu14,PDR14} that are processed, in combination with the simultaneous DM measurements, by new ``wideband timing'' analyses.

Wideband timing methods are still in their infancy, and most of the issues arising in their development come from the set of DM measurements made with the wideband TOAs; a wideband dataset is not just a set of TOAs.
One short-term solution is to simply ignore the DM measurements for the time being.
For NANOGrav, the current state of affairs is such that dispersive corrections are made for each pulsar by observing in (at least) two relatively large frequency windows, measuring multi-frequency TOAs in each, and fitting a piecewise constant DM model (``DMX'') as part of the timing model.
Insofar as the dispersive delay between the centers of the two bands is larger than the sum of the dispersive delays across each band individually, not much information is lost by ignoring the DM measurements in the wideband scheme.
This is not the case for all of NANOGrav's observational program, and so some sensitivity would be lost, not just from ignoring the DM measurements, but also from using TOAs from epochs separated by several days to infer the DM model \citep{Lam15}.
However, by simply disregarding the DM portion of the dataset, one can apply all of the latest timing analyses, noise modeling, and gravitational wave searches to the wideband TOAs as though they were conventional TOAs.

NANOGrav's multiband observational program highlights another issue.
As mentioned in Section~\ref{subsec:avg-port}, the absolute alignment of a particular band's portrait (and, thus, the model) will be wrong, and so the absolute DMs measured in disparate bands are expected to disagree by typically no more than an amount corresponding to about one phase bin of dispersive delay across the band; for our example data, this is an offset in absolute DM of $\sim$1$\times 10^{-3}$~pc~cm$^{-3}$ \citep[cf. Figure~4.5 of][]{PennucciPhDT}.
These DM offsets are analogous in the wideband scheme to the conventional TOA problem of aligning two templates from disparate frequencies -- one uses arbitrary phase JUMP parameters in timing analyses to minimize the residuals.
The offsets are very small relative to the absolute DM, but they have to be accounted for if the DM measurements are to inform the timing or DM model via DMX or otherwise\footnote{For more details on this subject see Section~4.2.2 of \citet{PennucciPhDT}.}.
For example, the difference in dispersive delay correction for a TOA at 820~MHz compared to one at 1.5~GHz that arises from a DM difference of 1$\times 10^{-3}$~cm~pc$^{-3}$ is $\sim$4~$\mu$s, a non-negligible amount.
Although these offsets can be readily measured and ``corrected'', they are better left as free parameters in a timing model, like JUMPs; analogous ``DMJUMP'' parameters, however, have yet to be anywhere implemented.

Relatedly, the noise modeling of TOA residuals has progressed significantly in the past few years \citep{Ellis13,vH13,vH14,Arzoumanian14,vH15,Ellis16}; similar -- and simultaneous -- noise modeling of the DM data will also be necessary to maximize sensitivity to gravitational waves.
Exactly how to make these advancements is a subject of ongoing study within NANOGrav and elsewhere.

\subsection{Future Directions}
\label{subsec:future}

There are several interesting avenues along which our modeling technique could be developed, most of which will be aided by the advent of very large fractional bandwidth systems.

First, accurate models of the expected pulse shape can be used to automatically detect changes in real-time or post-processing pipelines.
The most obvious use is for the flagging and zapping of channels contaminated by transient RFI.
A simple version of this is currently being used in the preparation of NANOGrav's 12.5-year wideband dataset.
More complicated programs, such as those using trained neural networks, could segregate various kinds of deviations from the expected pulse shape in order to detect small but significant changes, perhaps of astrophysical origin.

A second extension of our technique would be to combine it with Matrix Template Matching (MTM) \citep{vS06}.
MTM takes the original TOA phase-gradient algorithm \citep{Taylor92} and generalizes it to utilize all of the Stokes profiles.
A natural thing to do would be to model profile evolution of the average Stokes profiles using our methods and use them in an MTM algorithm to measure a TOA, the DM, and the rotation measure.
The importance of polarization calibration in precision timing experiments is becoming more apparent as PTAs dig further into the noise \citep{Dai15,Gentile18}, and this suggested combination could offer some insight.

Yet another extension could identify secular pulse profile changes by combining the principal component analysis of profile evolution parameterized by frequency with a similar decomposition parameterized by time.
This would look like a slow modulation of the profile evolution curve (e.g., Figure~\ref{fig:examp-proj}), and the aim would be to parametrize this profile evolution ``surface'' by frequency and time.
One application of such a development would be to track profile evolution in frequency and time of a precessing pulsar, thus improving the long-term timing of e.g. the double pulsar \citep{Kramer06}.

Fourth, as mentioned in Section~\ref{subsec:pca}, one drawback of the method is that it does not disentangle intrinsic profile evolution from e.g., pulse broadening from scattering in the ISM.
However, simulations could shed light on how arbitrary evolution curves from our models respond when their profiles are convolved with common pulse broadening functions, like a truncated one-sided exponential.
A profile evolution model disentangled from scattering could be used in a more general wideband TOA measurement algorithm that estimates a TOA, DM, and scattering parameters (Pennucci et al. (in prep.)).
High cadence observations of most pulsars at relatively low frequency (600~MHz) and relatively large fractional bandwidth (0.66) will soon be carried out by CHIME \citep{Ng17}; similar programs are currently being conducted at even lower frequency telescopes, such as the ongoing High-Band Antenna (HBA) observations (120--240~MHz) with the Low-Frequency Array (LOFAR) \citep{Stappers11}.
It would be interesting to investigate the temporal variability of scattering parameters in this way, and not just in the context of PTA experiments.

Finally, and relatedly, a first-principles study of why the profile evolution curves assume the shapes they do could be interesting from a theoretical standpoint.

We conclude with a statement that our proposed modeling technique is broadly applicable to a variety of pulsar observations, not just those used in PTA experiments.
Nevertheless, the results from modeling and timing all of the MSPs contained in NANOGrav's forthcoming 12.5-year dataset will be presented elsewhere.

Our publicly available code, ``PulsePortraiture''\footnote{\url{https://github.com/pennucci/PulsePortraiture}}, is readily usable on \texttt{PSRFITS} formatted archives in conjunction with the \texttt{python} interface of \texttt{PSRCHIVE} \citep{Hotan04}.
The modules \texttt{ppalign.py} and \texttt{ppzap.py} are relevant to Section~\ref{subsec:avg-port}, the module \texttt{ppspline.py} is relevant to Sections~\ref{subsec:pca}--\ref{subsec:model-const}, and the module \texttt{pptoas.py} is used in the creation of wideband TOAs.

\acknowledgements
\makeatletter{}The NANOGrav project receives support from National Science Foundation (NSF) Physics Frontiers Center award number 1430284.
The Green Bank Observatory is a facility of the National Science Foundation operated under cooperative agreement by Associated Universities, Inc.
TTP is supported by the Extragalactic Astrophysics Research Group (P.I. Zsolt Frei) funded by the Hungarian Academy of Sciences (Magyar Tudom\'{a}nyos Akad\'{e}mia).
TTP acknowledges and thanks the rest of NANOGrav's Timing Group for collecting and reducing the data presented in this paper, and Scott Ransom in particular for continued access to his computing cluster \texttt{nimrod}.
TTP would also like to thank KOI Creative Space, in which part of this manuscript was written, as well as Anya Bilous, Paul Demorest, Michael Lam, Aditya Parthasarathy, and the anonymous reviewer for their comments and ideas.

\software{PulsePortraiture \citep{PPsoftware,PPzenodo}, PSRCHIVE \citep{Hotan06}, PyWavelets \citep{pywt}, Matplotlib \citep{matplotlib}, SciPy \citep{scipy}, NumPy \citep{numpy}.}

\facilities{GBT (GUPPI)}
 
\hspace{0.25in}
\bibliographystyle{aasjournal}

\end{document}